\documentclass[twocolumn,preprintnumbers,aps,pra,superscriptaddress,amsmath,amssymb]{revtex4-1}

\usepackage{graphicx}
\usepackage{dcolumn}
\usepackage{bm}
\usepackage{bbm}
\usepackage{hyperref}
\usepackage{color}
\usepackage{multirow}
\usepackage{physics}

\usepackage{natbib}

\newcommand{\Ham}{\hat{H}}
\newcommand{\Prop}{\hat{U}}

\newcommand{\id}{\hat{\mathbb{I}}}

\newcommand{\sigz}{\hat{\sigma}_z}

\newcommand{\Sx}{\hat{S}_x}

\newcommand{\Sz}{\hat{S}_z}
\newcommand{\Sbf}{\hat{\mathbf{S}}}

\newcommand{\Iz}{\hat{I}_z}
\newcommand{\Ibf}{\hat{\mathbf{I}}}

\usepackage{braket}

\begin{document}

\title{Dynamical decoupling protocols with nuclear spin state selectivity} 

\author{J. E. Lang}
\email{jacob.lang.14@ucl.ac.uk}
\affiliation{Department of Physics and Astronomy, University College London, Gower Street, London WC1E 6BT, United Kingdom}

\author{J.-P. Tetienne} 
\affiliation{School of Physics, The University of Melbourne, VIC 3010, Australia}	

\author{T. S. Monteiro}
\affiliation{Department of Physics and Astronomy, University College London, Gower Street, London WC1E 6BT, United Kingdom}

\date{\today}
	
\begin{abstract}

The ability to initialise nuclear spins, which are typically in a mixed state even at low temperature, is a key requirement of many protocols used in quantum computing and simulations as well as in magnetic resonance spectroscopy and imaging. Yet, it remains a challenging task that typically involves complex and inefficient protocols, limiting the fidelity of ensuing operations or the measurement sensitivity. We introduce here a class of dynamical nuclear spin state selective (DNSS) protocols which, when applied to a polarised electron spin such as the nitrogen-vacancy (NV) centre in diamond, permit the addressing of selected nuclear states of the mixture. It works by splitting the underlying eigenstates into two distinct symmetries dependent on the nuclear spin state, and independent of the electron-nuclear coupling strength. As a particular example, we show that DNSS is achievable by simply introducing a detuning in the common Carr-Purcell-Meiboom-Gill (CPMG) protocol, where the state selection is then controlled by the inter-pulse spacing. This approach offers advantages in ultra-high fidelity initialisation of nuclear registers, ensemble polarisation and single-gate manipulation of nuclei.

\end{abstract}

\maketitle

The optical addressability and long room-temperature coherence times of electron spins of atomic defects, such as the 
ubiquitous NV$^-$ center in diamond \cite{Doherty2013}, are highly advantageous for  quantum sensing \cite{suter2016single, schirhagl2014nitrogen, Rondin2014, degen2017quantum} and computing \cite{childress2006coherent, ladd2010quantum, taminiau2014universal}. 
Dynamical decoupling pulse protocols protect a qubit state by averaging out the dephasing due to noise  \cite{hahn1950spin, carr1954effects, meiboom1958modified, maudsley1986modified,gullion1990new}. When the dynamical decoupling pulses are applied resonantly with a nuclear spin signal the decoupling fails and characteristic dips appear in coherence traces \cite{zhao2012sensing, kolkowitz2012sensing, taminiau2012detection}. The position and depth of these dips are used to extract information about the incident signal.
This has enabled detection of single nuclear spins and spin clusters inside the diamond \cite{zhao2012sensing, kolkowitz2012sensing, taminiau2012detection, zhao2011atomic, shi2014sensing}, ensembles of nuclear spins on the diamond surface \cite{staudacher2013nuclear, mamin2013nanoscale, rugar2015proton} and ultimately single nuclear spins on the diamond surface \cite{muller2014nuclear, sushkov2014magnetic, lovchinsky2016nuclear}. DD is also utilised in protocols for increasing spectral resolution \cite{boss2017quantum, schmitt2017submillihertz, laraoui2013high, zaiser2016enhancing, kong2015towards}.

\begin{figure}[t!]
\begin{center}
\includegraphics[width=2.5in]{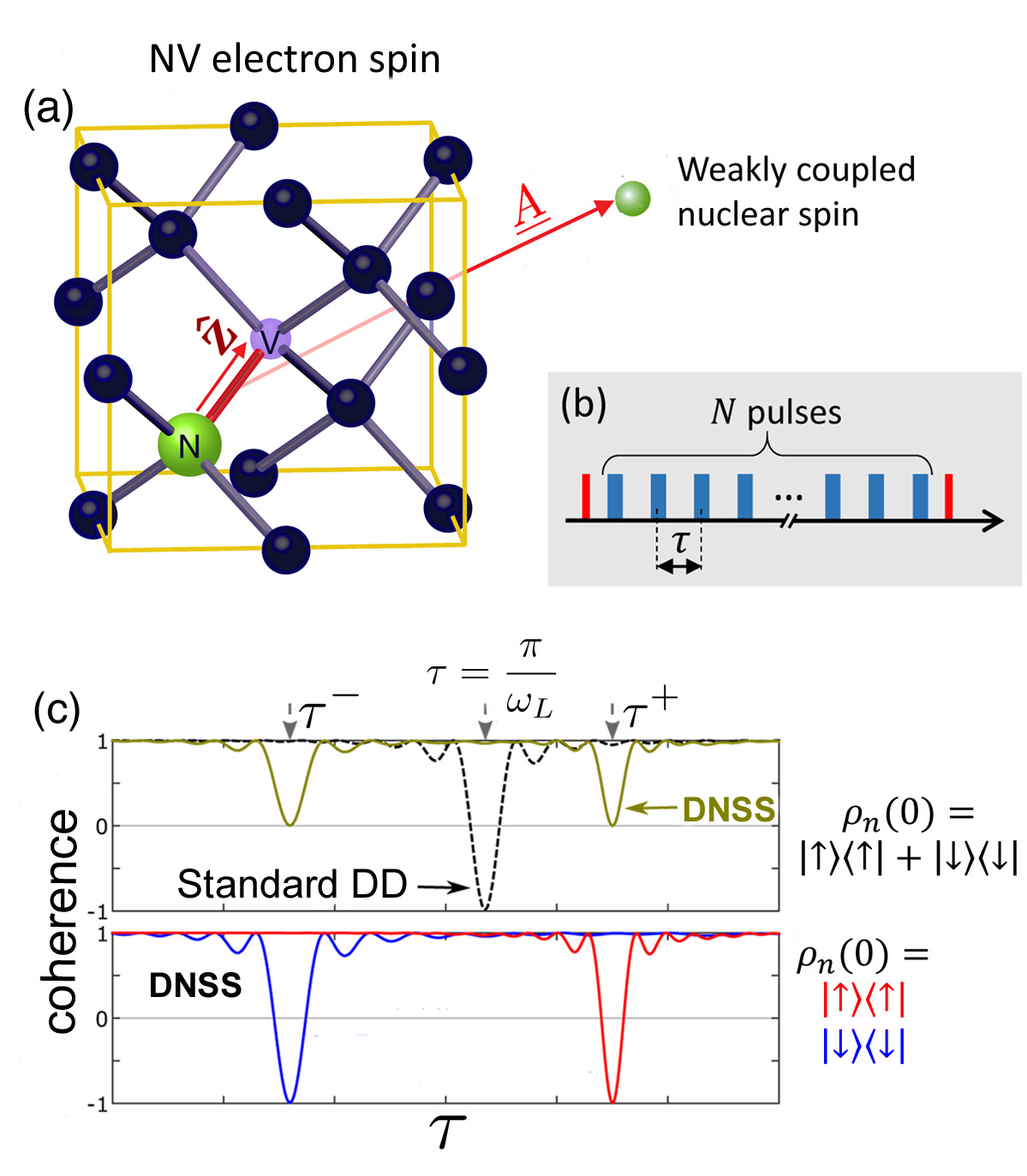}
\caption{{DNSS schematic \bf (a)} An electronic spin qubit, such as the NV spin, may interact with an external nuclear spin 
by means of a sequence of $N$ dynamical decoupling (DD) microwave pulses shown in
{\bf (b)}. {\bf (c)} For standard DD sequences, a  dip in the electronic spin qubit coherence is observed at  characteristic pulse spacing $\tau_{dip}\simeq \pi/\omega_L$ when weakly coupled to a nuclear spin. The nuclear spin is in an unpolarised mixed state $\rho_n$. For a DNSS sequence, {\em two} dips occur, at $\tau^\pm$. One dip addresses exclusively the `up' component of $\rho_n$, while the other addresses the `down' component, allowing high-fidelity manipulation and polarisation of nuclei. Plots correspond to a DNSS  sequence applied when the nuclear spin is initially mixed (top panel) or pure (bottom panel). The $\tau^\pm$ dip structure was recently observed experimentally \cite{PRA2018} but its remarkable  nuclear state selectivity  was, to date, overlooked. 
\\*}
\label{Fig1}
\end{center}
\end{figure}

\begin{figure*}[t!]
\begin{center}
\includegraphics[width=0.9\textwidth]{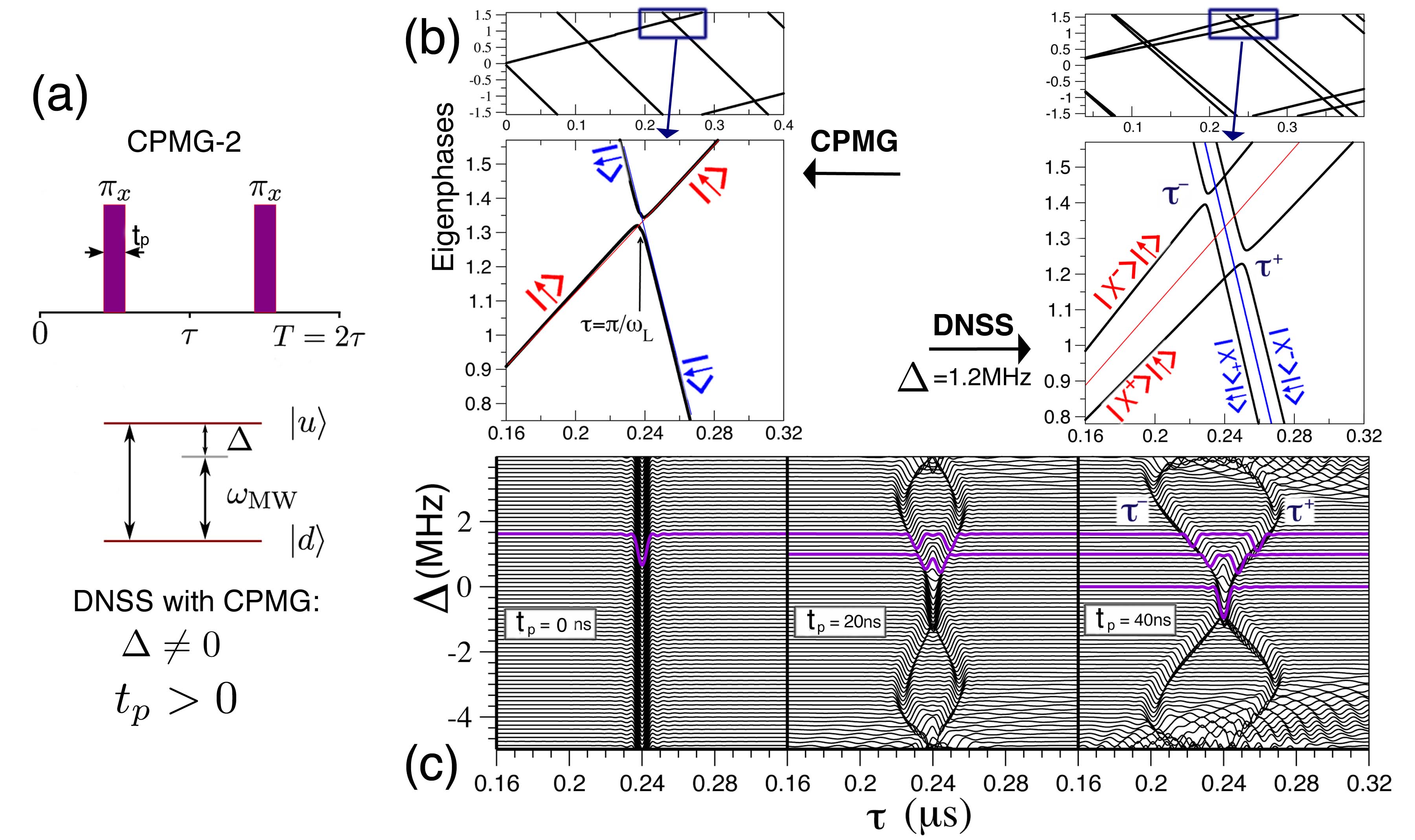}
\caption{{\bf (a)} An example of a DNSS sequence combines standard CPMG-2$N_p$  with a  pulse of effective duration $t_p>0$ and a non-zero detuning $\Delta$. CPMG-2$N_p$ corresponds to repeated application of $N_p$ pairs of pulses spaced by $\tau$.
{\bf (b)} In  dynamical decoupling (DD) a dip in electron spin coherence due to coupling to a nuclear spin corresponds to an avoided crossing of the 
underlying Floquet eigenstates \cite{Lang2015} . At this coherence dip,  nuclear and electronic states  become strongly correlated so this pulse spacing $\tau\sim\pi/\omega_L$ is used for  both sensing and manipulation of nuclear qubit registers. An initial $\pi/2$ pulse initialises the NV electronic spin state in a superposition $|X^+\rangle =\frac{1}{\sqrt{2}} (|u\rangle +|d\rangle)$ where $|u \rangle\equiv |m_s=\pm 1\rangle$ and $\ket{d} \equiv \ket{m_s = 0}$. The nuclear state is a mixture 
$\rho_n=\frac{1}{2}[ |\uparrow\rangle \langle\uparrow|+|\downarrow\rangle \langle\downarrow | ] $.
For  CPMG-2$N_p$  combining non-zero detuning $\Delta$ with $t_p>0$, the eigenstates now divide into two independent pairs of avoided crossings.  Importantly, the $ |X^+\rangle|\uparrow\rangle$ and $|X^+\rangle|\downarrow\rangle$ components of the initial mixture belong to separate branches (right panel) so the avoided crossings at $\tau^\pm$ each pick up {\em only one }  of the components of the mixture of nuclear spin-up or spin-down states. Hence this corresponds to DNSS. 
{\bf (c)} Calculated coherence traces as a function of detuning $\Delta$, for three different values of $t_p$. The parameters correspond to the experiments in Ref. [28], where a single NV electron spin at $\omega=1.5$~GHz interacts with an ensemble of proton spins with a Larmor frequency $\omega_L=2.1$~MHz. The number of pulses is $2N_p=336$. For the correspondence, here $A_\perp = 44$ kHz. The coherence traces in the right panel, $t_p=40$ ns, correspond closely to the measured 
traces  shown in Fig.3(a) of \cite{PRA2018}.
The location of $\tau^\pm(\Delta,t_p)$  is essentially independent of the NV-nuclear couplings $A_\perp$, which affect only the dip height and width, not their central value.  Note that the $X$ symmetry set by the pulse is key: the effect is eliminated by eg XYN sequences that do not preserve this symmetry.  }
\label{Fig2}
\end{center}
\end{figure*}

The potential of nuclear spins as a memory resource on account of their long-lived coherence properties 
is well-established. Recently they have been used as registers, including in demonstrations of error correction  with NV qubits subject to DD control \cite{childress2006coherent, ladd2010quantum, taminiau2014universal}. In addition,
nuclear registers have allowed nanoscale NMR for biosensing with unparalleled sensitivity \cite{Aslam2017}. 
A recent review  of the applications of auxiliary nuclear spin registers is given in \cite{Awschalom2018}.
The difficulty here is that while NV electronic spins may be initialised optically,  the nuclear spin states 
 cannot and are typically in a thermally mixed state; DD based gates rotate all components of the mixture simultaneously and initialisation  involves a sequence of gates interspersed with timed free evolution, limiting efficiency and fidelity. Recent methods for initialising the state of nuclear spins include lab-frame and rotating frame resonance matching \cite{broadway2018quantum, London2013} -- which relies on delicate control of the microwave or static magnetic field strength -- and combinations of different DD sequences \cite{schwartz2018robust, taminiau2014universal}.

Here we introduce a new protocol which selects individual components  of the nuclear mixture, \emph{in a single gate}, allowing for drastically enhanced control of the nuclear spin registers. It also offers new possibilities for dynamical nuclear polarization, of practical significance for enhancing NMR sensitivity. These protocols achieve dynamical nuclear spin-state selectivity (DNSS) through a slight modification of common DD protocols such as CPMG, which breaks the symmetry between the different nuclear spin states. Below, we present the physical mechanism underlying DNSS, which is a fully quantum effect.

\begin{figure*}[t!]
\begin{center}
\includegraphics[width=0.8\textwidth]{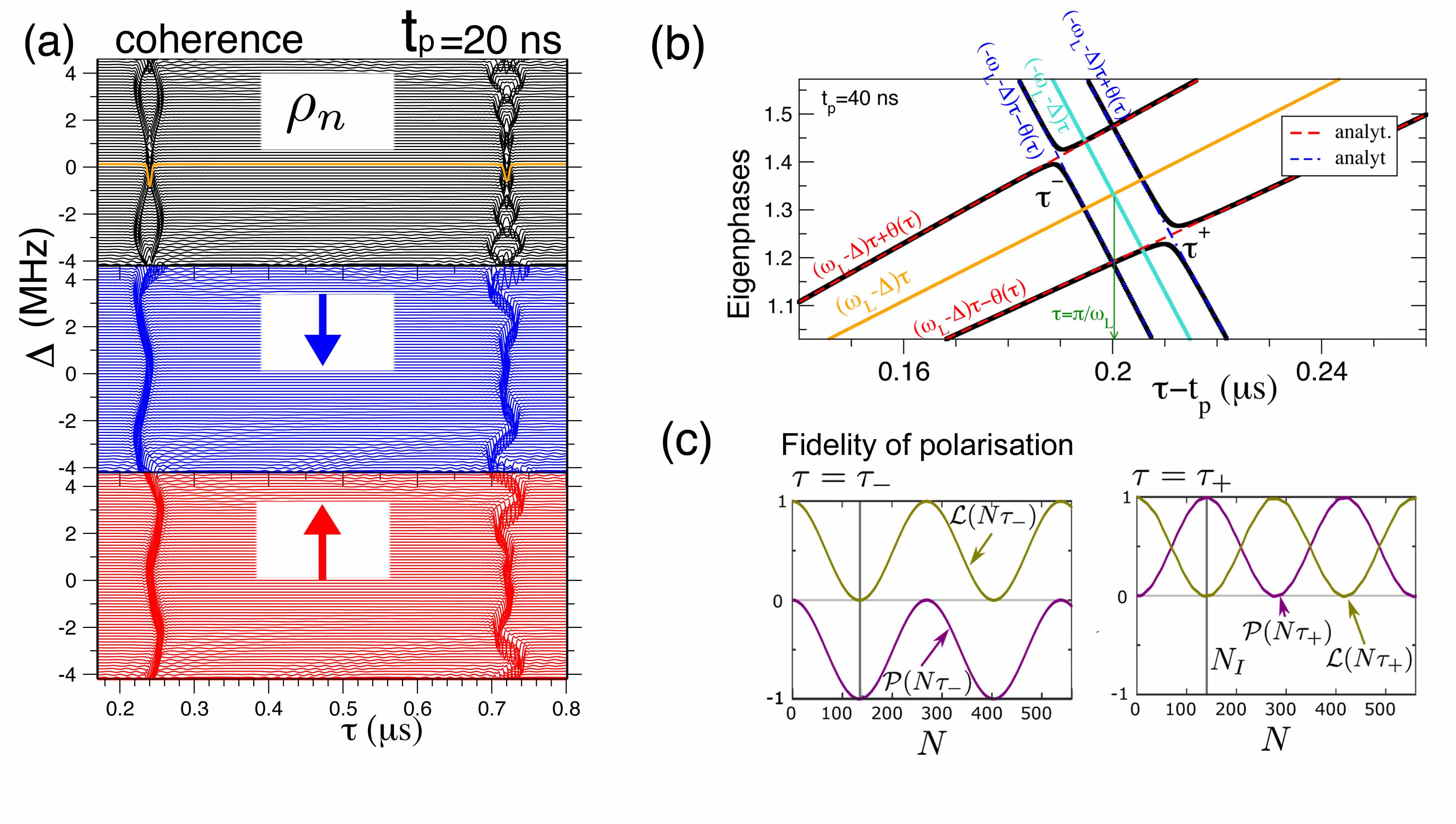}
\caption{{\bf (a)} Coherence traces calculated with the same parameters as in Fig.~\ref{Fig2}(b) but for a wider range of times $\tau$ to capture the first two harmonics, at $\tau\simeq\pi/\omega_\text{L}$ and $\tau\simeq 3\pi/\omega_\text{L}$. The upper panel 
shows a calculation with the incoherent average over  the full initial mixed state $|X^+\rangle\otimes \rho_n$ ( the NV spin is first initialised into $0\rangle$, then a $\pi/2$ pulse rotates it to one of $|X^\pm\rangle$ states; here we choose 
$|X^+\rangle$). 
Below we show the same calculation, but including only either the `up' or `down' components of $\rho_n$, showing that each 
corresponds to a single one of the two coherence dips. Initialising with $|X^-\rangle$ would give the opposite behavior.
{\bf (b)} Comparison between the numerics and analytical calculation that obtains the $t_p$-dependent corrections to the Floquet phases $\pm \theta(\tau)$. Note that a single eigenphase $\theta(\tau)$, obtained from  $\Prop_p(T)$ corrects all four Floquet phases
and determines both $\tau^+$ and $\tau^-$.
{\bf (c)} Shows that we can use this technique to initialise and control with high fidelity a  more weakly coupled nuclear spin of $^{13}$C.  Simulated here is the polarisation of a single nucleus $\mathcal{P}(N\tau) = \braket{2\Iz}$, compared with the coherence $\mathcal{L}(N\tau) = \braket{2\Sx}$ of the ($m_s = 0, m_s = -1)$ NV qubit. We take $B_z = 400$~G and $\{A_x, A_z\}/2\pi = \{10, 0\}$~kHz. A pulse number scan of the coherence and nuclear polarisation for the above parameters but with fixed pulse spacing $\tau = \tau^\pm$. The number of pulses required to initialise the nuclear spin is denoted by $N_I$ (grey vertical lines).}
\label{Fig3}
\end{center}
\end{figure*}
We consider the Hamiltonian, 
\begin{equation}
\Ham(t) = \omega_\text{L}\Iz + \Sz\textbf{A}\cdot\Ibf + \Ham_p(t)
\label{Ham}
\end{equation}
for a general electron spin qubit ($\Sbf$) coupled to a single nuclear spin qubit ($\Ibf$) and subject to CPMG control
illustrated in Fig.~\ref{Fig1}(a), where magnetic field is aligned along the $z$-axis and we make the usual pure-dephasing approximation. $\omega_\text{L}$ is the nuclear Larmor frequency; the hyperfine field  $\textbf{A}$ felt by the nuclear spin has components
 ${A_\perp, A_\parallel}$ relative to the $z$-axis and $\Ham_p(t) = \Delta\Sz + \Omega(t)\Sx$ is the pulse control Hamiltonian. $\Omega(t)$ is the microwave drive strength which is non-zero during the pulses and $\Delta$ is the microwave detuning from resonance. For DD control using the CPMG sequence, the microwave pulses are applied along the $x$-axis at regular intervals, $\tau$, as shown in Fig.~\ref{Fig1}.  For top-hat pulses in the absence of detuning, a perfect $\pi$-rotation is obtained when the pulse height $\Omega = \pi/t_p$ for a pulse width of $t_p$ (though we note DNSS may be achieved for any pulse shape). For a nuclear spin precession characterised by the Larmor frequency, a coherence dip 
appears at characteristic pulse spacing $\tau = \pi/\omega_\text{L}$. For the NV center the nuclear Larmor frequency contains a small shift due to the parallel component of the hyperfine coupling but for generality we  include this in the Larmor frequency, $\omega_\text{L} \pm A_\parallel/2 \Rightarrow \omega_\text{L}$.

While state propagation under ${\hat H}(t)$ enables simulation of DNSS, here we apply a quantum analysis to explain the underlying physical mechanism. As the Hamiltonian is periodic, $\Ham(t + T) = \Ham(t)$, Floquet theory provides the natural framework for analysing the dynamics. We seek the Floquet modes, $\ket{\Phi_l^F}$ - the eigenstates of the one-period evolution operator
 \begin{equation}
{\hat U}(T)|\Phi^F_l\rangle = \lambda_l| \Phi^F_l\rangle \equiv \exp(-i({\mathcal E}_l(\tau) + k2\pi) )   |\Phi^F_l\rangle
\label{Floquet}
\end{equation}
where the eigenvalues of the $l$-th  eigenstate are  $\exp(-i({\mathcal E}_l(\tau) + k2\pi) )$  and where the $+k2\pi, \forall k \in \mathbb{Z}$ arises from the temporal periodicity of $\hat{H} (t)$. The  ${\mathcal E}_l+ 2\pi k$ for $k=0,1,2...$ are the system \emph{Floquet phases} which are simply the eigenphases of the one-period evolution operator. In  \cite{Lang2015} it was shown that coherence dips in DD {\em all} correspond to avoided crossings of Floquet phases as illustrated in Fig.\ref{Fig2}(b).

The positions of the avoided crossings are revealed by the true crossings in the unperturbed Floquet phase spectrum (see Fig.~\ref{Fig3}(b)). The unperturbed spectrum is found by setting $\textbf{A} = 0$. In this case $\Ham^{\textbf{A} = 0}(t) = \omega_\text{L}\Iz + \Ham_p(t)$ so the one-period evolution operator is given by $\Prop^{\textbf{A} = 0}(T) = \exp(-i\omega_\text{L}\Iz T)\Prop_p(T)$ where $\Prop_p(T)$ is the one-period pulse propagator. For simple pulse sequences, such as CPMG-2$N_p$, the pulse propagator can be constructed in a straightforward manner and for small detuning errors the pulse propagator takes the approximate form 
\begin{equation}
\Prop_p(T) \approx -\id + i 2\theta(\tau)\Sx \approx \exp(-i(\pi\id + 2\theta(\tau)\Sx)),
\end{equation}
see Supp. Info. The explicit form of the small quantity $\theta(\tau) \ll 1$ is determined by the type of pulse error and is given, for detuning errors and flip-angle errors, in the Supp. Info. For ideal CPMG control $\Prop_p(T) = -\id$ as $\theta = 0$.

The unperturbed Floquet spectrum is thus given by $\varepsilon(\tau) = \pi \pm \omega_\text{L}\tau \pm \theta(\tau) + k2\pi$ with the corresponding Floquet modes $\ket{X_\pm, \uparrow\downarrow}$. The dip positions, $\tau_\pm$, are then found by solving $\tau_\pm = (\pi \pm |\theta(\tau_\pm)|)/|\omega_\text{L}|$. (Note that the additional slope, $-\Delta\tau$, of the Floquet phases seen in the figures in due to the $\Delta/2$ energy shift in the NV spin operator $\Delta\Sz = \Delta/2(\sigz + \id)$. The additional slope does not affect the dip positions or the system dynamics.) $\theta(\tau)$ is an oscillatory function of $\Delta t_p$ and at points where $\theta(\tau)=0$, a single dip is restored. In Fig.~\ref{Fig3}(a) we see that for the fundamental signal, this occurs at higher detunings, $\Delta\simeq 4$ MHz,
but subsequent harmonics yield several values of  $\theta(\tau)=0$. As was illustrated in Fig.\ref{Fig2}(c)) for 
$\Delta\gtrsim 4$ MHz, the approximation $\Delta/2 <<\Omega$ fails and the $\tau^\pm$ structure breaks ups.

The corresponding Floquet modes are labelled in Fig.~\ref{Fig2}(b). Importantly for DNSS based on CPMG-2, $\Prop_p(T)$ has eigenstates of the same symmetry $|X_\pm\rangle= \frac{1}{\sqrt{2}} [|u\rangle \pm |d\rangle ]$. Thus, away from avoided crossings, the eigenstates of the system, to an excellent approximation correspond to $|\Phi^F_{\pm, \uparrow}\rangle=|X_\pm\rangle |\uparrow\rangle$ and $|\Phi^F_{\pm, \downarrow}\rangle=|X_\pm\rangle |\downarrow\rangle$. 

Fig.\ref{Fig2}(b) (right panels) shows that instead of the single avoided crossing seen for $t_p=0$, we now have 4 possible crossings near $\tau=\pi/\omega_L$. Whether they are open - and produce an experimental dip -  or closed (and are thus true crossings which give no signal) depends on the coupling $\hat{S}_z\textbf{A}\cdot\Ibf$.  Since  $\langle X^\pm |\hat{S}_z| X^\mp \rangle \langle \downarrow| \hat{I}_x | \uparrow \rangle \neq 0$ we see that  these correspond to a pair 
of distinct avoided crossings which give the two separate coherence dips as shown in Fig.\ref{Fig2}(c), as a function of $t_p$ and $\Delta$, illustrated with a nuclear spin with $\omega_\text{L} =  2.1$ MHz for which $A_\perp=44$ kHz.  However we note that two crossings between the two symmetry subspaces shown (shown also in Fig.\ref{Fig3}(b))  are true crossings, since $\langle X^\pm| \hat{S}_z|X^\pm\rangle=0$. Both true crossings are at $\tau=\pi/\omega_L$.

Fig.\ref{Fig3} illustrates the effect of nuclear spin selectivity.  The NV electronic spin is prepared in the state $|X^+\rangle\otimes \rho_n$, where $\rho_n$ represents the $ \uparrow, \downarrow$ nuclear mixed state.  
Due to the structure of $\hat{U}(T)$, the $|X^+\rangle |\uparrow \rangle$ and $|X^+\rangle| \downarrow \rangle$ are in this case remarkably close to one of two Floquet eigenstates. However these eigenstates each belong to a distinct avoided crossing; at $\tau^+$, the $|X^+\rangle |\uparrow \rangle$ rotates into $|X^-\rangle| \downarrow \rangle$, but cannot couple to the other states. The $|X^+\rangle |\downarrow \rangle$ components remain unaffected by the pulse sequence. 
Fig.\ref{Fig3}(a) represents a numerical simulation showing the clean separation between the contributions of the two nuclear spin states. To our knowledge, no previous single pulse sequence has been found to exhibit this 
spin-selective property. 

The DNSS protocol works by introducing a small $x$-rotation into the pulse propagator - where usually DD would be designed to produce an identity operator. Whilst this additional rotation would typically be considered erroneous it actually creates the DNSS effect itself. As long as the electron qubit is initialised into an initial $\ket{X_\pm}$ the DNSS effect can be included without significantly degrading the coherence time, as evidenced in previous work \cite{PRA2018}. Noting that DNSS is provided by a small $x$-rotation, an alternative DNSS protocol can be proposed by simply introducing a small additional rotation to each $\pi_x$ pulse in the CPMG sequence (and without any detuning). This may prove experimentally convenient. A similar eigenstate structure is obtained. Care is needed to preserve the refocussing properties of the DD which obtains $T_2 >>T_2^*$. In the case of the  \cite{PRA2018} experiments  with $N\sim 100-200$ the splitting was achieved without  loss of refocussing.  

Fig.\ref{Fig3} illustrates  DNSS for detection of more weakly coupled $^{13}$C.  The simulations in Fig.\ref{Fig3} (c)  show that a one-gate $\pi$ rotation of the nuclear state can be achieved with 0.9995 fidelity (discounting experimental errors). 
 After an $N\simeq 140$ DNSS sequence, an initial $|X^+\rangle\otimes \rho_n$ yields a mixed  $|X^\pm\rangle$ for the electronic spin qubit but a fully polarised nuclear state. At the point where $\mathcal{L}(N\tau) = \braket{2\Sx}=0$ (also illustrated in Fig.\ref{Fig1}) the mixed state $\rho_n$ has been swapped onto the NV. However an NV state may be rapidly reinitialised optically without disturbing the nuclear state, so initialisation 
leaves both in a pure state.

{\em Conclusions} The essential ingredient for practical realisation of DNSS sequence is the introduction of a  controlled `error' in the pulse rotation angle of the DD sequence carefully calibrated to maintain its refocussing properties.
Given the  technological potential of NV spin qubits in diamond,  substantial effort  is being devoted by several groups worldwide to perfecting pulse  protocols that minimise errors. This may have led to DNSS possibilities being overlooked: the CPMG sequence is perceived to be less robust to pulse errors relative to the widely used  XY-N sequences; however, the latter,  by eliminating alternating pulse symmetries, eliminates the eigenstate splitting that underpins DNSS. In the present paper we show that, paradoxically,  
what was initially examined simply as a deleterious  error, actually enables a uniquely powerful new form of state  control.
The robustness of DNSS protocols to uncontrolled experimental errors is currently being investigated, which will ultimately determine its practicality for applications.

\bibliographystyle{apsrev4-1}
\bibliography{bib}

\end{document}